\documentstyle[epsf,twocolumn,aps]{revtex}
\newcommand{\eq}{\begin{equation}}
\newcommand{\ee}{\end{equation}}
\newcommand{\eqa}{\begin{eqnarray}}
\newcommand{\eea}{\end{eqnarray}}

\renewcommand{\v}[1]{{\bf #1}}

\newcommand{\s}{\sigma}
\newcommand{\pprl}{Phys. Rev. Lett.} 
\newcommand{\pprb}{Phys. Rev. {B}} 
\begin{document}
\twocolumn[
\hsize\textwidth\columnwidth\hsize\csname@twocolumnfalse\endcsname
\draft

\title{
Inhomogeneous d-wave superconducting state of
a doped Mott insulator
}

\author{Ziqiang Wang, Jan R. Engelbrecht, Shancai Wang, Hong Ding}
\address{Department of Physics, Boston College, Chestnut Hill, MA 02467}
\author{Shuheng H. Pan}
\address{Department of Physics, Boston University, Boston, MA 02215}
\date{\today}
\maketitle

\begin{abstract}
Recent atomic resolution scanning tunneling microscope (STM) measurements
discovered remarkable electronic inhomogeneity, i.e. nano-scale spatial
variations of the local density of states (LDOS)
and the superconducting energy gap,
in the high-T$_c$ superconductor Bi$_2$Sr$_2$CaCu$_2$O$_{8+x}$.  Based on the
experimental findings, we conjectured that the inhomogeneity arises from
variations in local oxygen doping level and may be generic of doped Mott 
insulators.
In this paper, we provide theoretical support for this picture. We study 
a doped Mott insulator within a generalized t-J model, where doping is 
accompanied by ionic Coulomb potentials centered in the BiO plane located 
a distance $d_s$ away from the CuO$_{2}$ plane.  We solve, at the mean-field 
level, a set of spatially unrestricted Bogoliubov-de Gennes equations 
self-consistently to obtain the distributions of the hole concentration, 
the valence bond and the pairing order parameters for different 
nominal/average doping concentrations.  We calculate the LDOS spectrum, the
integrated LDOS, and the local superconducting gap as those measured by STM,
make detailed comparisons to experiments, and find remarkable agreement with
the experimental data.  We emphasize the unconventional screening of the ionic
potential in a doped Mott insulator and show that nonlinear screening dominates
on nano-meter scales, comparable to the short coherence length of the 
superconductor, which is the origin of the electronic inhomogeneity.  It
leads to strong inhomogeneous redistribution of the local hole density and
promotes the notion of {\it local doping concentration} (LDC).
We find that the inhomogeneity structure manifests itself at all
energy scales in the STM tunneling differential conductance, and elucidate the
similarity and the differences between the data obtained in the constant
tunneling current mode and the same data normalized to reflect constant
tip-to-sample distance.  We also discuss the underdoped case where nonlinear
screening of the ionic potential turns the spatial electronic structure into a
percolative mixture of patches with smaller pairing gaps embedded in a
background with larger gaps to single particle excitations.

\end{abstract}
]

\section{Introduction}

The parent compounds of the high-T$_c$ superconductors are stoichiometric
Mott insulators \cite{anderson}. 
In principle, carriers can be introduced into the
insulating state either by field effect 
or by chemical doping, 
but to date high-$T_c$ superconductivity in the cuprates has only been 
found to arise when they are properly doped away from stoichiometry.
In the case of YBa$_2$Cu$_3$O$_{7-x}$ (YBCO) and Bi$_2$Sr$_2$CaCu$_2$O$_{8+x}$
(BSCCO), the dopants are the excess oxygen atoms.
The latter inject holes into the CuO$_2$
planes, leaving behind the negatively charged oxygen ions.
While for YBCO, the dopant ions go into the copper oxide chains and can
be arranged more periodically, \cite{liang}
in the case of BSCCO, the dopants, with
a concentration of $x$, disorder themselves in the semiconducting BiO layer 
which is about $5$\AA \ away from the CuO$_2$ plane where superconductivity
is believed to originate.

In addition to doping holes into
the CuO$_2$ planes, the dopant oxygen atoms inevitably introduce 
long-range random ionic potentials that
scatter the carriers in the copper-oxide plane in analogy to the 
situation encountered in two-dimensional (2D) electron systems in
modulation doped semiconductors. 
This has the potential
to cause the electronic structure in the copper-oxide plane to be 
inhomogeneous.
Early neutron scattering\cite{neutron}, tunneling\cite{tunneling}
and STM\cite{earlystm} data 
showed features that could be accounted for by inhomogeneity, but they
were usually attributed to sample quality.
The common view, which is largely based
on the physics of ordinary metals, has been that the ionic potential
in a high-quality crystal would be screened by phonons and 
by the carriers in the plane 
such that charges distribute uniformly and are accommodated by a 
homogeneous electronic structure; albeit that the latter has the tendency 
towards microscopic phase separation\cite{kivelson}
due to strong electronic correlations.
This viewpoint appears to be supported, to a certain extent, 
by transport\cite{transport} and
photoemission\cite{photoemission} measurements
on BSCCO which observe
quasiparticles with relatively well-defined
momentum, a mean-free path of about $100$\AA \ in the nodal 
direction, and a transport mean-free path of about $400$\AA.

However, this conventional view has been seriously challenged by recent 
low-temperature STM measurements on BSCCO.\cite{pan,howald,davis} 
In Ref.~\cite{pan}, Pan et al. (here 
after referred to as we)
observed spatially cross-correlated variations in the LDOS and
the superconducting energy gap on a remarkably short length scale
of about $14$\AA. We identified this as the gap amplitude coherence
length or the pair size. It directly confirms the local pairing
nature in the high-T$_c$ superconductor. The magnitude of the
superconducting gap has a Gaussian distribution with
a mean value of 42 meV, close to the gap value obtained from
planar tunneling,\cite{tunnelinggap}
and a full-width-half-maximum of about 20 meV. 
This variance coincides with the 
intrinsic width ($\sim$ 20 meV) of the coherence peak measured by
angle resolved photoemission spectroscopic (ARPES) near the 
antinodes where the d-wave gap is at its maximum. \cite{hong} It offers an 
explanation of the latter as arising, even in the absence of
bilayer splitting, from averaging over
a distribution of local gap values under the macroscopic
ARPES light spot.

Careful analysis of the STM data, combined with those of the ARPES,
led us to conjecture in Ref.~\cite{pan} that the observed
inhomogeneous electronic structure arises from the ionic potential 
associated with the off-stoichiometry oxygen dopants 
disordered in the BiO layer. We argued that the screening of the
ionic potential in the doped Mott insulator is unconventional and incomplete on
the nano-meter scale of the pair size, 
causing the local doping (hole) concentration
to exhibit spatial variations. Such a doping 
profile results in a highly cross-correlated pairing gap profile
if the pairing interaction is sufficiently short-ranged.

It should be stressed that the physics involved here is highly unconventional.
In an ordinary metal such as copper, the Thomas-Fermi
screening length in the linear screening theory
is shorter than $1$\AA, beyond which the
ionic potential is perfectly screened and the charge density homogeneous.
Such a short screening length is a result of the large
thermodynamic density of states or the compressibility
($dn/d\mu$) in ordinary metals. However, adding holes to a doped Mott 
insulator causes significant shift in the chemical potential due
to the strong local Coulomb repulsion. This leads to a small electronic
compressibility and thus a larger screening length even within the
linear screening theory.
We believe that the observed inhomogeneous
superconducting state is a result of the
strong local correlation in a doped Mott insulator.

It is important to provide theoretical tests for these 
ideas and to compare theoretical results with experiments.
In this paper, we do so by studying the d-wave superconducting state
of a doped Mott insulator in a microscopic theory of a
generalized t-J model,
where doping is accompanied by ionic Coulomb potentials  
centered a distance $d_s$ away from the 2D plane.
Since an exact solution of this model for large systems is not possible even
numerically at present, we 
provide a self-consistent mean-field solution that captures
the essential physics. The hope is that upon comparing the results
to experiment, we can develop useful insights to guide our
understanding in the interim. Specifically, we solve a
set of spatially unrestricted Bogoliubov-de Gennes equations
self-consistently at every site on square lattices of up to
$32\times32$ sites to obtain the spatial distributions of
the hole concentration, the resonating valence
bond and the pairing order parameters for different nominal/average
doping concentrations. 
We find that the inhomogeneity originates from
nonlinear screening of the ionic potential.
We show that a single negatively charged test ion
inserted at a setback distance $d_s$ above the 2D plane induces
a nonlinear screening cloud wherein the doping concentration
is significantly larger than its averaged value. For a given strength
of the Coulomb interaction, the size of the
nonlinear screening cloud is controlled by the distance $d_s$
and is {\it independent} of the average doping concentration.
With a finite density of dopant ions
(same as the averaged doped hole density),
an inhomogeneous electronic structure
on the scale of $d_{s}$ emerges that shares an analogy to the 
2D electron system in modulation doped semiconductors where nonlinear
screening leads to an inhomogeneous mixture of metallic and dielectric
regions at low electron density.\cite{efros}
The immediate consequence of the
nonlinear screening is the strong spatially inhomogeneous redistribution
of the charge carrier density brought about by the bare ionic
potential. The concept of a {\it local doping concentration} (LDC)
\cite{pan} therefore
naturally arises. The d-wave pairing amplitude turns out to follow locally
the spatially varying hole density,
resulting in an inhomogeneous superconducting state.
In the resonating valence bond picture\cite{anderson},
this corresponds to an inhomogeneous state of spinon pairing and
local holon condensation. 
We calculate the local tunneling density of states
spectrum, the integrated LDOS, and the local superconducting gap
as those measured by STM. Surprisingly, the
local spectral properties can be well described by
the Mott-Hubbard picture once the notion of a
LDC is established.
We perform a detailed comparison of our results with the experimental data
and find remarkable agreement in support of the conjectures made in
Ref.~\cite{pan} and the theoretical picture of 
spinon pairing with local holon condensation for the 
inhomogeneous superconducting state. Extending this picture to
the underdoped regime, where the averaged inter-hole separation becomes
larger than the setback distance $d_s$, naturally leads to
percolative structures of superconducting patches immersed in
a background with a large tunneling gap to single particle excitations.

Currently, there exists a ``discrepancy'' in the published 
STM data by different groups\cite{pan,howald,davis} with regard to whether
the LDOS shows spatial inhomogeneity at low energies.
We point out that the difference in the presented experimental data 
depends on whether they have been normalized to remove 
certain matrix element effects.
In Ref.~\cite{pan}, we have taken into account, by correctly normalizing
the data, the effect of the tunneling matrix element along the
direction perpendicular to the surface.
The normalized data reflect the tunneling spectra when
the tip to sample distance is kept effectively at a constant. 
We believe this is physical. 
The low energy electronic spectrum as measured by the local tunneling 
density of states is indeed inhomogeneous. \cite{pan} 
In this paper, we show that
the same inhomogeneity structure manifests itself at {\it all} energy scales
in the calculated tunneling spectra. We find that the calculated
zero-bias tunneling conductance, which is 
dominated by the nodal quasiparticles, shows spatial variations
that are correlated with that of the LDC.
We elucidate in detail the mapping, the differences and similarities
between the tunneling differential conductance
obtained in the constant tunneling current mode, which does not
directly represent the electron LDOS, and the same data normalized
to reflect a constant tip-to-sample distance that faithfully represent
the electronic tunneling density of states.
We also provide preliminary evidence that the high superconducting
transition temperature $T_c$ can be protected by
the short superconducting coherence length and coexist with
this type of electronic inhomogeneity.

The rest of the paper is organized as follows. In section II, we
describe our generalized t-J model and set up the self-consistent
equations for the spatially unrestricted Hartree-Fock-Bogoliubov
mean-field solutions. In section III, we
study the screening properties of the ionic potential in detail.
A single test ion is inserted which imposes a coulombic potential 
on the otherwise uniform d-wave superconducting state. 
The charge redistribution and the response of the pairing order parameter
to screening of the ionic potential is analyzed.
In section IV, we discuss the solution of the inhomogeneous d-wave
superconducting state for a finite density of dopant ions and
doped holes. We emphasize the concept of LDC
and analyze the distribution of the
d-wave order parameter and its correlation with the
local hole concentration in subsection IV.A. 
The numerical results 
of the local tunneling spectrum are presented in subsection IV.B
and compared to experimental
data. We analyze the statistical properties of the LDC,
the integrated LDOS and the pairing
gap distribution and the spatial correlation and cross-correlation
among these physical observables.
In section V, we focus on the inhomogeneity at low energy scales.
We study the spatial distribution of the zero-bias tunneling conductance,
its correlation with the LDC, and make
predictions for experimental tests. In section VI, we discuss the
mapping between constant-current differential tunneling conductance
and the constant tip-to-sample distance LDOS spectra. The STM topography
is calculated for our system together with the constant-current differential
tunneling conductance and compared to experiments.
A summary of the results and their implications are given in section VII
together with discussions of several open issues.

\section{Generalized $\rm\lowercase{t}-J$ model 
and the self-consistency equations}

We begin with the generalized t-J model Hamiltonian that includes the
long-range Coulomb interaction and the ionic potentials introduced by
the process of carrier doping,
\eq
H=H_{t-J}+ H_{\rm Coul}+H_{\rm ion}.
\label{h}
\ee
Here $H_{t-J}$ is the usual 2D t-J model on a square lattice,
\eq
H=-t\sum_{\langle i,j\rangle}(c_{i\s}^\dagger c_{j\s}+{\rm h.c.})
+J\sum_{\langle i,j\rangle}(\v S_i\cdot\v S_j\!-{1\over4}n_i n_j),
\label{htj}
\ee
where $c_{i\s}^\dagger$ is the electron creation operator and 
$\v S_i$ is the spin operator $\v S_i=({1\over2})
c_{i\alpha}^\dagger{\vec \sigma}_{\alpha\beta}c_{i\beta}$.
The sums over $\langle i,j\rangle$ are among nearest neighbors and sums
over repeated spin indices are implied.
The most important Mott-Hubbard physics, i.e. the strong on-site Coulomb
repulsion, is included in the additional constraint of no double occupancy 
at each site, $n_i=c_{i\s}^\dagger c_{i\s}\le1$.
$H_{\rm Coul}$ is the long-range Coulomb repulsion,
\eq
H_{\rm Coul}=\sum_i V_in_i,\quad V_i=V_c\sum_{j\ne i} {n_j-{\bar n}\over\vert
\v r_i-\v r_j\vert},
\label{hcoul}
\ee
where ${\bar n}$ denotes the average density. The Coulomb 
interaction strength $V_c$ is given by 
$V_c=e^2/4\pi\epsilon a$, with $\epsilon\simeq8$ the
dielectric constant\cite{hyberstein} and $a\sim3.8$ \AA \  
the lattice constant corresponding to the Cu-Cu atomic spacing.
The resulting $V_c\sim t$. As emphasized in Ref.~\cite{dhlee1},
the inclusion of the long-range Coulomb interaction is
necessary to prevent the mean-field ground state of $H_{t-J}$ 
from macroscopic phase separation in the interesting parameter 
regime.\cite{note1} The off-plane ionic potential of the dopants is
described by $H_{\rm ion}$,
\eq
H_{\rm ion}=\sum_i U_i n_i, \quad U_i= \sum_{{\rm ion}=1}^{N_{\rm ion}}
{V_{\rm ion}\over\sqrt{\vert \v r_i-\v r_{\rm ion}\vert^2+d_s^2}}.
\label{ionpotential}
\ee
Here $d_s$ is the distance between the CuO$_2$ plane and
the BiO layer where the negatively charged ions reside randomly
at $\v r_{\rm ion}$,
and $N_{\rm ion}$ is the number of independent ions in the BiO layer.
To model the situation in BSCCO, where each dopant oxygen gives
one hole to each of the planes in a bi-layer, we use
$N_{\rm ion}=N_{\rm hole}=x\times N_s$ where
$N_{\rm hole}$ is the number of doped holes, $x$ is the {\it average}
doping (hole)
concentration on a lattice of $N_s$ sites.
We use $J$ as the unit of energy and
set $d_s=1.5 a$, $V_{\rm ion}=V_c=5 J$, and $t=3J$ in
most of our numerical calculations. We verified that varying these
parameter in a reasonable range does not qualitatively 
change our results.

It is convenient to describe the  
projected Hilbert space in terms of a spin-carrying fermion, the spinon
$f_{i\s}^\dagger$, creating the singly occupied site with spin-$\s$
and a spinless boson, the holon $b_i$, keeping track 
of the empty site \cite{kotliarliu}. The electron creation operator 
becomes $c_{i\s}^\dagger=f_{i\s}^\dagger b_i$ and the
occupancy constraint translates in this slave-boson formulation  
into $f_{i\s}^\dagger f_{i\s}+b_i^\dagger b_i=1$.
In the mean-field theory, the antiferromagnetic spin-exchange
term is decoupled according to\cite{ubbenslee}
\eqa
{\v S_i\cdot\v S_j} = &-& {3\over8}[\chi_{ij}^*f_{i\s}^\dagger f_{j\s}
+{\rm h.c.} \nonumber \cr
& +& \Delta_{ij}^*(f_{i \downarrow} f_{j \uparrow}
-f_{i \uparrow}f_{j \downarrow})
+{\rm h.c.}]
\nonumber \cr
&+& {3\over8}(\vert\chi_{ij}\vert^2+\vert\Delta_{ij}\vert^2),
\label{ss}
\eea
where $\Delta_{ij}$ and $\chi_{ij}$ are the spinon pairing and bond
order parameters respectively,
\eq
\Delta_{ij}=\langle f_{i \downarrow} f_{j \uparrow}
-f_{i \uparrow}f_{j \downarrow}\rangle,\quad
\chi_{ij} = \langle f_{i\s}^\dagger f_{j\s}\rangle,
\label{deltachi}
\ee
defined for each nearest neighbor bond.
The inhomogeneous superconducting phase is reached through
the local condensation of bosons at low temperatures,
\eq
\langle b_i^\dagger \rangle =\langle b_i \rangle
= {\bar b}_i.
\label{b}
\ee
It is important to emphasize that in the presence of the translation
symmetry breaking ionic potential in Eq.~(\ref{ionpotential}), 
$(\Delta_{ij},\chi_{ij},{\bar b}_i)$ become spatially dependent and must
be determined at every site self-consistently.
Note that the local doped hole density or the LDC
is then directly related to the local boson condensate density,
\eq
x_i={\bar b}_i^2,
\label{xi}
\ee
and the constraint at the mean-field level becomes
$
n_{i}^f=f_{i\s}^\dagger f_{i\s}=1-x_i,
$
which is enforced on average at every site by locally shifting the
fermion chemical potential $\mu_f$ to
$\lambda_i+\mu_f$. Throughout this paper, we use $x$ to denote the
average (doped) hole density or the nominal doping concentration,
\eq
x={1\over N_s}\sum_{i=1}^{N_s} x_i.
\label{x}
\ee

The decoupled Hamiltonian can be written down using the 
Bogoliubov-Nambu formalism over the Hilbert space of paired spinons,
\eq
H=\sum_{i,j}(f_{i\uparrow}^\dagger,f_{i\downarrow})
\pmatrix{K_{ij} & F_{ij}\cr
F_{ji}^* & -K_{ji}^*\cr}
\pmatrix{f_{j\uparrow}\cr f_{j\downarrow}^\dagger},
\label{mfh}
\ee
where the sums on $i$ and $j$ run over all lattice sites and
\eqa
F_{ij}&=&{3\over8}J\Delta_{ij}\sum_\eta\delta_{j,i+\eta},
\label{fij} \\
K_{ij}&=&-(tb_i^2+{3\over8}J\chi_{ij})\sum_\eta\delta_{j,i+\eta}
+[V_{\rm sc}(i)-\mu_f]\delta_{ij}, 
\label{kij}
\eea
with $\eta=\pm{\hat x},\pm{\hat  y}$. In Eq.~(\ref{kij}), $V_{\rm sc}(i)$
given by
\eq
V_{\rm sc}(i)= U_i+\lambda_i+V_c\sum_{j\ne i}{{\bar b}_j^2-x\over
\vert \v r_i-\v r_j\vert},
\label{vsc}
\ee
is the nonlinearly screened local potential seen by the spinons
implied by the self-consistency conditions. 
Note that in deriving Eq.~(\ref{vsc}),
only the Hartree potential in
the long-range Coulomb interaction is retained, whereas 
the exchange potential is neglected. The effects of the latter on
the superconducting state will be studied elsewhere.

The Hamiltonian in Eq.~(\ref{mfh}) can be diagonalized in real space
by solving the corresponding Bogoliubov-de Gennes equations to obtain
the eigenstates $\gamma_n^\dagger$ and $\gamma_n$ with energy
$E_n, n=1,\dots,2N_s$. The spinon operator can be expanded
in this basis according to
\eqa
f_{i\uparrow}^\dagger(t)&=&\sum_n u_n(i)\gamma_n^\dagger e^{-iE_nt/\hbar}
\label{upu} \\
f_{i\downarrow}(t)&=& \sum_n v_n(i)\gamma_n^\dagger e^{-iE_nt/\hbar},
\label{downv}
\eea
where $(u_n(i),v_n(i))$ is the wave-function at site $i$. The order
parameters and the local hole density can be expressed in terms
of the wave-functions as follows
\eqa
\Delta_{ij}=\sum_n \bigl[v_n^*(i) u_n(j)[1&-&f(E_n)]
\nonumber \\
&-&u_n(i) v_n^*(j)f(E_n)\bigr],
\label{dijuv} \\
\chi_{ij}=\sum_n \bigl[ v_n^*(i)v_n(j)[1&-&f(E_n)]
\nonumber \\
&+&u_n(i)u_n^*(j)f(E_n) \bigr],
\label{chiijuv} \\
1-x_i=
\sum_n \bigl [\vert v_n(i)\vert^2 [1&-&f(E_n)]
\nonumber \\
&+&\vert u_n(i)\vert^2 f(E_n)\bigr],
\label{holedensity}
\eea
where $f(E_n)$ is the usual Fermi distribution function.

We solve the self-consistency Eqs.~(\ref{mfh}-\ref{holedensity}) through
numerical iterations.
Typically, we start with a random set of $(\Delta_{ij},
\chi_{ij},\lambda_i,x_i,\mu_f)$, insert them into Eq.~(\ref{mfh}) and
diagonalize the resulting matrix to obtain the wave-functions 
$(u_n(i),v_n(i))$ and the eigenvalues $E_n$. Then we update the
set of $(\Delta_{ij},\chi_{ij},\lambda_i,x_i,\mu_f)$ according
to Eqs.~(\ref{dijuv},\ref{chiijuv},\ref{holedensity}) and insert them back into
the Hamiltonian (\ref{mfh}). The procedure is iterated until 
convergence is reached. In general we allow $\chi_{ij}$ and 
$\Delta_{ij}$ to be complex at the beginning of the iteration,
but both of these order parameters converge to real
values at the end of the iteration in the average doping
range ($0.06< x < 0.32$) studied.

In order to compare to STM data, we calculate 
the local {\it tunneling} density of states,
\eq
N_i(\omega)={\rm Im}\int dt e^{i\omega t} G_{ii}^{\rm ret}(t),
\label{nig}
\ee
where $G_{ii}^{\rm ret}(t)$ is the {\it retarded} local Green's function 
of the {\it electrons},
\eq
G_{ii}^{\rm ret}(t)=-\theta(t)\langle\{c_{i\s}^\dagger(0),c_{i\s}(t)
\}\rangle.
\label{g}
\ee
In the mean-field theory, since the bosons locally condense, we have
\eq
G_{ii}^{\rm ret}(t)
=-{\bar b}_i^2\theta(t)\langle\{f_{i\s}^\dagger(0),f_{i\s}(t)\}\rangle.
\label{gret}
\ee
Notice that the local electron Green's function and therefore the local
tunneling density of states is equal to the local holon condensate
density times the local spinon Green's function.\cite{wang} From
Eq.~(\ref{xi}), it is clear then that the LDOS is overall proportional to
the LDC $x_i$. Using Eqs.~(\ref{upu},\ref{downv}),
we obtain,
\eq
N_i(\omega)=x_i\sum_n\bigl[\vert u_n(i)\vert^2\delta(\omega-E_n)
+\vert v_n(i)\vert^2\delta(\omega+E_n)\bigr] .
\label{ldos}
\ee
The role played by the LDC is two-fold: it
enters as an overall prefactor through essentially the wave-function
renormalization of the electrons as well as through the 
LDOS of spinons from the response of $\Delta_{ij}$ and $\chi_{ij}$
to a spatially varying $x_i$. 
This is a property of the tunneling density of states which
is different from the thermodynamic
density of states.\cite{elihu}

Our choice of the t-J model for quantitative calculations is a natural one, 
because it is the simplest model that captures the physics of 
a doped (antiferromagnetic) Mott-insulator.
In this strong coupling approach, the carrier concentration is 
proportional to the doping concentration (rather than the electron density),
which is the most fundamental property of a doped Mott-insulator.
More specifically, as can be clearly seen from Eq.~(\ref{gret}),
the coherent weight of the quasiparticle is small
at small doping and scales with the LDC. This is 
in excellent agreement with the recent ARPES measurements
\cite{hong,shen} on BSCCO, where the weight of the emergent
quasiparticle peak below $T_c$ near the anti-node was found to be 
proportional to the doping concentration at low temperatures.

\begin{figure}   
\center   
\centerline{\epsfxsize=2.8in   
\epsfbox{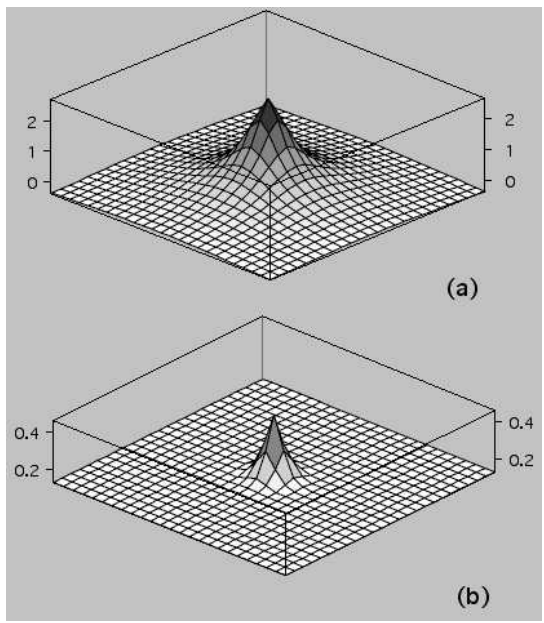}
}
\begin{minipage}[t]{8.1cm}   
\caption{The Coulomb potential from a single negatively charged ion located
at a distance $d_s=1.5a$ above the center of an otherwise
uniform 2D d-wave superconductor. The lattice size is $25\times25$ and
the average doping $x=0.12$. (a) The long-range bare ionic potential
from Eq.~(\ref{ionpotential}).
(b) The screened potential from Eq.~(\ref{vsc}).
} 
\label{oneionvimp}    
\end{minipage}   
\end{figure}   

\section{A single off-plane ion and nonlinear screening}

To understand the screening properties of the ionic potential better, 
we start with
the case where a single off-plane ion (i.e. $N_{\rm ion}=1$) 
is placed above the
center of a $25\times25$ lattice with a setback distance $d_s=1.5a$.
The ion imposes a Coulomb potential
on the otherwise uniform superconducting state at a doping level $x$.
This situation is similar to that of a single nonmagnetic impurity (say Zn)
extensively studied using the t-J model recently,
\cite{sigrist,tsuchiura,zhu} except that the impurity
is located out of the plane and the impurity potential is long-ranged.
The scattering is presumably {\it not} in the unitary limit.
In addition, fully self-consistent solution in (${\bar b}_i$, $\Delta_{ij}$,
$\chi_{ij}$) has not been studied before.

Fig.~1 shows the bare ionic potential $U_{i}$ from Eq.~(\ref{ionpotential})
and the screened potential $V_{\rm sc}(i)$ from Eq.~(\ref{vsc}) coming 
out of the self-consistent solution. The average doping concentration 
in the plane is $x=0.12$. It is clear that the long-range
impurity potential in Fig.~1a is
perfectly screened at distances much larger than $d_s$ but poorly screened
at short distances resulting in a short-range impurity
potential in Fig.~1b. 
\begin{figure}   
\center   
\centerline{\epsfxsize=2.8in   
\epsfbox{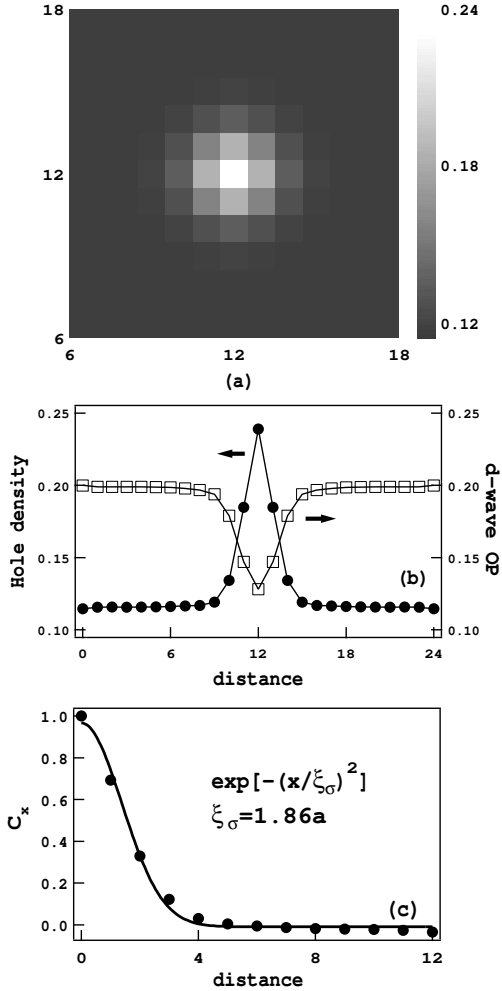}
}
\begin{minipage}[t]{8.1cm}   
\caption{Spatial variations of the LDC induced by the
ionic potential in Fig.~1.
(a) The 2D plot of the local hole density ($x_i$) 
redistribution as a result of nonlinear screening.
(b) The hole density and the corresponding d-wave pairing order parameter
(OP) along a horizontal line-cut passing through the center of the
$25\times25$ lattice at $x=0.12$. (c) The azimuthally averaged 
auto-correlation function of the spatial variations in local hole
concentration $\delta x_i$ defined in Eq.~(\ref{xcorr}). The solid line
is a Gaussian fit to the data, which gives a decay length 
$\xi_\s=1.86a$.
} 
\label{oneionholedensity}    
\end{minipage}   
\end{figure}   

The nature of the screening is revealed
when we look at the redistribution of the carrier density as
a result of screening. The self-consistently determined
local hole concentration $x_i$ is plotted in a 2D hole density map 
in Fig.~2a, which shows that
in the vicinity of the dopant ion, the hole density is strongly
inhomogeneous, namely the changes in the hole concentration $\delta x_i$,
brought about by the dopant ionic potential, are comparable to the averaged
concentration $x$ itself. Fig.~2b shows the hole density 
distribution along a horizontal
line-cut passing through the center of the lattice.
This clearly demonstrates that the screening of
the ionic impurity potential in the doped Mott insulator is
highly unconventional and is entirely
dominated by nonlinear screening\cite{efros} on the scale of the
ionic setback distance $d_s$. The nonlinearality in screening is further
verified by its nonlinear response to changes in the strength
of the Coulomb potential $V_{\rm ion}$.
It is instructive to compare to conventional screening in the linear
response theory described
by the Thomas-Fermi screening length, which is applicable to
ordinary metals and superconductors. For an electron gas in 2D,
the Thomas-Fermi screening length is 
{\it independent of electron density} and is given by
$q_{TF}^{-1}=a_B^*/2$, where $a_B^*=\hbar^2\epsilon/m^*e^2$ is 
the effective Bohr radius.\cite{ando} Taking $\epsilon\simeq8$ and a
thermodynamic effective mass $m*/m\sim3$, which is appropriate for 
the cuprates, gives $a_B^*\simeq1.4$\AA.
The resulting Thomas-Fermi screening length $\sim0.7$\AA \  which
is much less than a lattice spacing $a\simeq3.8\AA$. Thus linear
screening theory would have predicted that the ionic Coulomb potential
is screened on the length scale of a lattice spacing and the charge
redistribution induced by the bare potential is small.
In our case, Figs.~2a and 2b clearly
show that linear screening is only valid when the distance to the
center of the ionic potential is much greater than the distance $d_s$,
where the hole density is essentially uniform and its
value close to the average concentration $x=0.12$. 

Thus, the profile of the screened impurity potential and that of the hole
density distribution and, as we shall see later, that of the d-wave
pairing OP (see Fig.~2b), are controlled 
by the crossover from nonlinear to linear screening as one moves away 
from the center of the ionic potential.
To better quantify this crossover behavior, we study the auto-correlation
function of the local hole density variations $\delta x_i=x_i-x$,
\eq
{ C}_x(\v r_j)=
{1\over N_s}\sum_i\langle \delta x_i\delta x_{i+j} \rangle.
\label{xcorr}
\ee
In Fig.~2c, we show the azimuthally averaged ${C}_x$,
which decays very fast with the distance, indicative of strong short-range
correlation in the LDC.
The short-distance behavior can be fitted very well by a Gaussian 
with a decay length $\xi_\s=1.86 a$. For convenience, let us define
the correlation length by the decay length of the Gaussian function.
We arrive at a hole-density correlation length $\xi_x=1.86a$, which is
quite close to the setback distance $d_s=1.5a$, suggesting the latter
as the length scale over which the crossover between linear and
nonlinear screening takes place.
\begin{figure}   
\center   
\centerline{\epsfxsize=2.6in   
\epsfbox{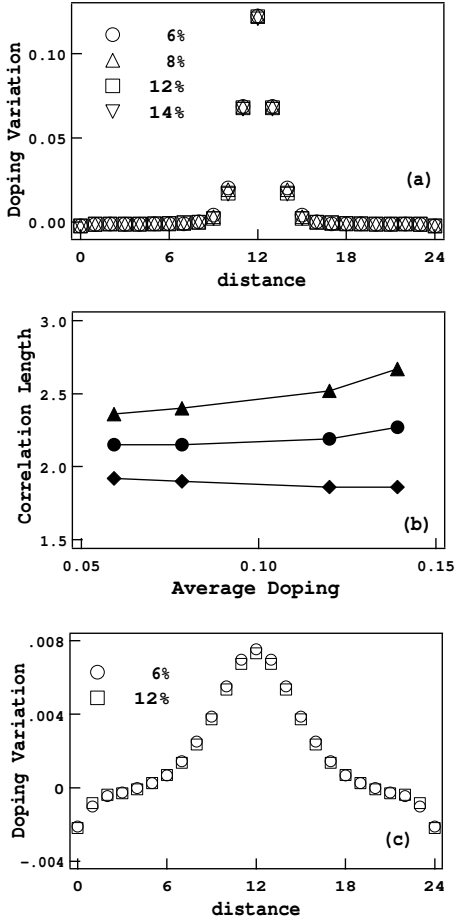}
}
\begin{minipage}[t]{8.1cm}   
\caption{The average doping and setback distance dependence of
screening. (a) The spatial variations in hole concentration
($\delta x_i$) along the central line-cut at four different
average doping concentrations, showing screening is insensitive
to average doping $x$.
(b) The decay lengths of the hole-density (diamonds)
and the d-wave pairing OP (triangle) auto-correlations, and that of the
cross-correlation (circle) between the two as a function of 
average doping $x$.
(c) Same as in (a), but for a larger setback distance $d_s=5a$.
} 
\label{oneiondopingdep}    
\end{minipage}   
\end{figure}   
We next study the average doping dependence of the screening in
the nonlinear regime. In the linear screening theory, the fact that
the 2D Thomas-Fermi screening length is independent of carrier density
is a consequence of the properties of the density of states and 
the polarizability of the 2D electron system.\cite{ando}
This suggests that the insensitivity of screening to carrier
concentration in 2D should be true even for nonlinear screening.
Analytical solutions of the 2D nonlinear screening problem only
exists for the special case of an antidot in a semiconductor
2D electron gas in the continuum.\cite{antidot} 
To study this question in the d-wave superconducting state
of a doped Mott insulator in the presence of a periodic lattice potential,
we plot, in Fig.~3a, the variation of the local hole density from
its average value ($\delta x_i$) as a function of distance along
the center horizontal cut for several different average hole
concentrations. The fact that the data points almost completely
collapse onto a single curve shows that the screening is 
indeed insensitive to the carrier density in both the linear and 
the nonlinear screening regime. This is verified quantitatively
by the very weak doping dependence of the decay lengths
extracted from the Gaussian fits to the correlation functions at
the corresponding doping levels, shown in Fig.~3c.  
Thus, for a given strength of the ionic Coulomb potential, the size of the 
nonlinear screening cloud around the center of the potential 
is controlled entirely by the setback distance $d_s$. 
In Fig.~3c, we show the spatial variation of the LDC
$\delta x_i$ along
the center horizontal cut for the case of a larger setback distance
$d_s=5a$ at $x=0.06$ and $0.12$.
It is clear that as the screened potential becomes smoother, the size of
the screening cloud becomes larger, characterized by a decay length on 
the order of $d_s$. 
\begin{figure}   
\center   
\centerline{\epsfxsize=3.8in   
\epsfbox{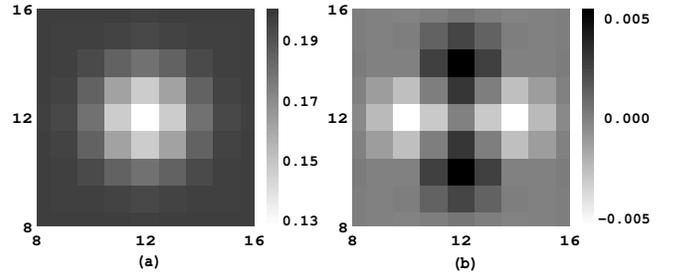}
}
\begin{minipage}[t]{8.1cm}   
\caption{The 2D maps showing the spatial variations in the
dominant d-wave (a), and the accompanying extended s-wave (b)
pairing OPs in the vicinity of the ionic potential.
} 
\label{oneionscop}    
\end{minipage}   
\end{figure}   
In Fig.~4, we show the spatial variation of the self-consistently
determined superconducting order parameter near the center of the
ionic potential for $d_s=1.5a$. 
We find that the nonlinear screening region where the hole 
concentration exhibits strong inhomogeneity is accompanied by
spatial variation of the order parameter $\Delta_{ij}$.
Moreover, when $\Delta_{ij}$, defined in Eq.~(\ref{deltachi}),
is decomposed into a d$_{x^2-y^2}$ and an extended
s-wave component at a site $i$ according to,
\eqa
\Delta_d(i)&=&{1\over4}(\Delta_{i,i+{\hat x}}+\Delta_{i,i-{\hat x}}
-\Delta_{i,i+{\hat y}}-\Delta_{i,i-{\hat y}}),
\label{deltadi} \\
\Delta_s(i)&=&{1\over4}(\Delta_{i,i+{\hat x}}+\Delta_{i,i-{\hat x}}
+\Delta_{i,i+{\hat y}}+\Delta_{i,i-{\hat y}}),
\label{deltasi} 
\eea
we find that the spatially varying d-wave order parameter $\Delta_d(i)$
shown in Fig.~4a is complimented by a much smaller nonuniform
$\Delta_s(i)$ shown in Fig.~4b. The magnitude of $\Delta_d$ along
the center line cut in Fig.~4a has been shown in Fig.~2b.

This situation is quite reminiscent
of a single {\it in-plane} nonmagnetic impurity, e.g. an intentionally doped
Zn atom, that replaces
a copper atom causing strong scattering in the unitary limit.
\cite{sigrist,tsuchiura,zhu,sasha,franz} 
The inhomogeneity induced $\Delta_s$ shows an
interesting four-fold (d-wave) symmetry, i.e. it is vanishingly small
along the nodal direction of $\Delta_d(i)$ and changes sign upon
a $90^\circ$ rotation. These properties can be
qualitatively understood from the Ginzburg-Landau 
theory\cite{sigrist,berlinsky} that
permits by symmetry a mixed gradient term proportional to
$\partial\Delta_d\partial\Delta_s$. Note that
the order parameters in this $d+s$ state are
both real and that spontaneous time-reversal symmetry breaking, 
such as in a $d+is$ state, does not occur. It is interesting
to point out that 
the reduction is greater in the anti-nodal direction where
the d-wave gap is at its maximum. 
\begin{figure}   
\center   
\centerline{\epsfxsize=3.6in   
\epsfbox{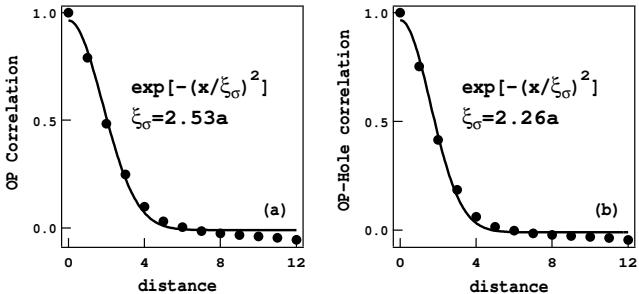}
}
\begin{minipage}[t]{8.1cm}   
\caption{The azimuthally averaged auto-correlation function
for the d-wave OP variation (a) and the {\it absolute} value of the
cross-correlation function between the hole density variation
and the d-wave OP variation (b). The solid lines are Gaussian fits
to the data.
} 
\label{oneioncorrelation}    
\end{minipage}   
\end{figure}   

The auto-correlation function of the spatial variation
of the d-wave OP $\delta\Delta_d(i)=\Delta_d(i)-{\bar\Delta_d}
$ and the cross-correlation function between 
$\delta\Delta_d(i)$ and the local doping variation $\delta x_i$
are defined by
\eqa
{C}_{\Delta_d}(\v r_j)&=&{1 \over N_s}\sum_i
\langle \delta\Delta_d(i)
\delta\Delta_d(i+j)\rangle,
\label{deltacorr} \\
{ C}_{x-\Delta_d}(\v r_j)
&=&{1\over N_s}\sum_i 
\langle \delta x_i\delta\Delta_d(i+j)\rangle.
\label{xdeltacorr}
\eea
We plot the azimuthally averaged correlation functions in Fig.~5a and 5b.
The $\Delta_d$-correlation length 
$\xi_{\Delta_d}\simeq 2.53a$, which is close to the hole density correlation
length. The fact that the cross-correlation length, $\xi_{x-\Delta_d}
\simeq 2.26a$, is close to $\xi_x$ and $\xi_{\Delta_d}$ shows
that the spatial variations in $\Delta_d$
and the LDC are strongly (anti-)correlated.
This is an important character of the d-wave superconducting
state emerged from the short-range resonating valence bond
state,\cite{anderson,kivelsonrvb} where the local spinon pairing
produced by the nearest-neighbor antiferromagnetic exchange interaction 
is directly affected by the local holon condensate density. 

To summarize this section, an out-of-plane negatively charged
ion induces a nonlinear screening cloud in its vicinity,
wherein the doping concentration is significantly higher and 
the d-wave pairing parameter significantly smaller than
their averaged values in the bulk of the 2D superconductor.
The size of the nonlinear screening cloud is essentially independent
of the averaged doping concentration. It is
controlled by the setback distance $d_s$ of the dopant ions to the
2D plane. This local picture for the response of the 
d-wave superconducting state
to a single ionic test charge will be important for understanding
the real situations of many ions, particularly in the case of dilute
dopants in the underdoped regime. In the rest of the paper, we
use $d_s=1.5a$. This comes out naturally from the lattice constants
of BSCCO which is about $3.8$\AA \ in the plane and $5$\AA \ in the
$c$-direction. We will come back to this issue in the last section
and discuss other physical processes, such as dielectric screening
from interband transitions and electron-phonon coupling,
that could complicate the choice of $d_s$, but would not change
the essential physics described in this paper.
\begin{figure}   
\center   
\centerline{\epsfxsize=3.0in   
\epsfbox{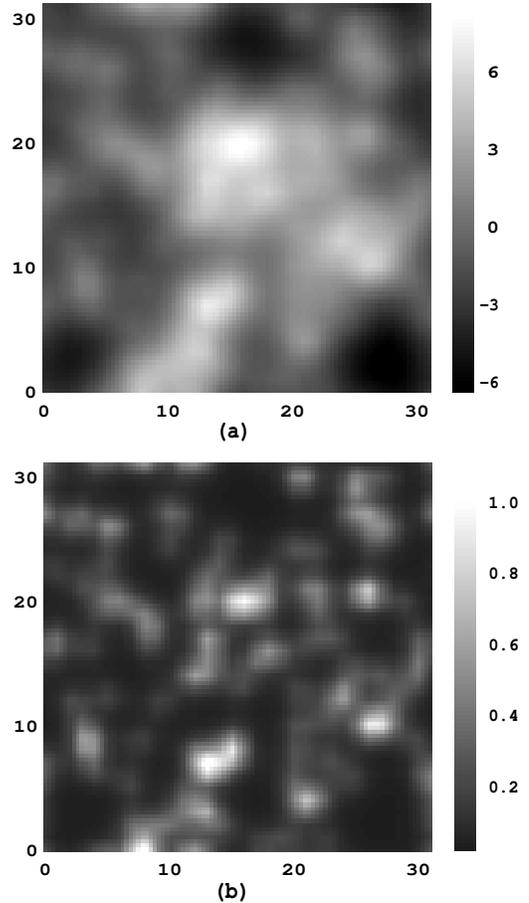}
}
\begin{minipage}[t]{8.1cm}   
\caption{2D density plot of the bare (a) and the screened (b) ionic
potential on a $32\times32$ lattice at average doping $x=0.12$. 
The data has been smoothed
to reduce the effects of the lattice discreteness.
} 
\label{vimpvscr}    
\end{minipage}   
\end{figure}   
\section{ Finite density of ions --- inhomogeneous superconducting state of
a doped Mott insulator}

Having studied how an external out-of-plane charge is screened by
the d-wave superconductor, we now turn to the situation
in BSCCO where $N_{\rm ion}$ number of negatively charged
ions resides in the BiO layer donating
$N_{\rm hole}=N_{\rm ion}$ number of doped holes to each of the 
CuO bilayers. Without the knowledge of the detailed chemical frustration
encountered by the dopant ions, we assume that they distribute
randomly in the BiO layer. This can lead to accidental clustering
of the dopant oxygen atoms which may or may not happen in real
materials depending on the details of the chemistry.
Since very strong clustering happens rarely and locally, it does not
affect the essential physics we describe.

\subsection{ Inhomogeneous electronic structures}

The bare ionic potential calculated from Eq.~(\ref{ionpotential})
is shown in Fig.~6a for a $32\times32$ lattice at $x=0.12$. 
The self-consistently determined screened potential of Eq.~(\ref{vsc}) 
is plotted in Fig.~6b. 
The large fluctuations in the bare ionic potential 
are dramatically screened as can be seen by comparing Fig.~6a and Fig.~6b.
The spatial variations of the self-consistently determined LDC
in Eq.~(\ref{xi}) and $\Delta_d$ in Eq.~(\ref{deltadi})
are shown as 2D maps in Figs.~7a and 7b. 
It is clearly seen from Fig.~7a that the
variations in the hole density closely track 
the screened potential in Fig.~6b. The local ionic dopant 
configurations are shown as black dots. The reverse color-coding for
the $\Delta_d$-map makes it easy to observe the (anti-)correlation between
$x_i$ and $\Delta_d(i)$.
The average inter-hole distance at this doping
is $d_x\simeq 2.86a$, which is smaller than the size of the screening
cloud $2\xi_x\simeq3.72a$ of a single isolated ion discussed in 
the previous section.
The fact that the number of dopant ions is the same as the number of
doped holes then implies that the screening clouds associated with
individual ions overlap and interfere such that nonlinear screening
dominates and leads to significant inhomogeneity
in the electronic structure throughout the entire system.
This confirms the conjecture we made in Ref.~\cite{pan} that the
inhomogeneity results from the lack of conventional metallic
screening of the dopant ionic potential in doped Mott insulators.
\begin{figure}   
\center   
\centerline{\epsfxsize=3.0in   
\epsfbox{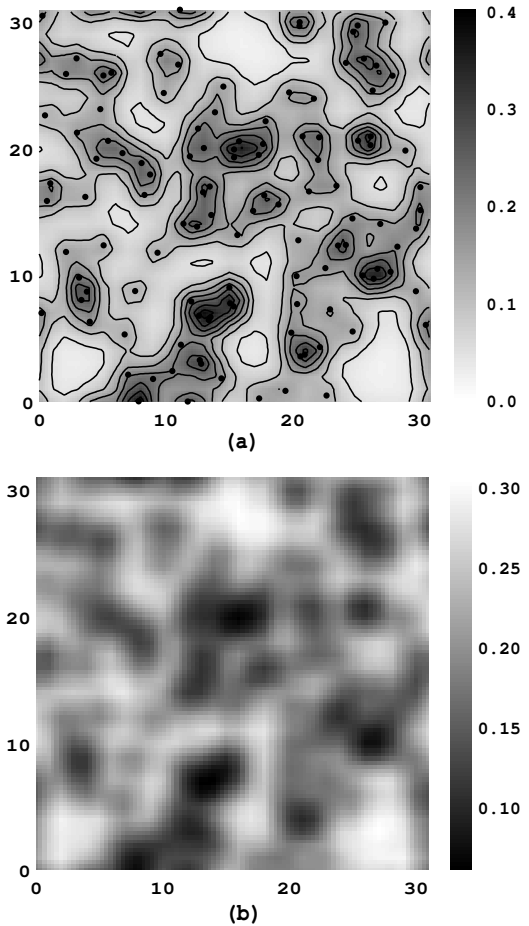}
}
\begin{minipage}[t]{8.1cm}   
\caption{The spatial variations of the local hole concentration
$x_i$ (a) and the d-wave order parameter $\Delta_d(i)$ (b)
shown as 2D maps for $x=0.12$ on a $32\times32$ lattice.
The location of the negatively charged ions projected on the
2D plane is shown as black dots in (a), together with the
contours of the doped hole density. Reversed color-coding is used
in (b) for easy visualization of the anti-correlation between
$x_i$ and $\Delta_d(i)$.
} 
\label{holemap}    
\end{minipage}   
\end{figure}   

We present, in Fig.~8, the auto- and cross-correlation function
analyses for the variations in LDC and the d-wave OP
defined in Eqs.~(\ref{xcorr},\ref{deltacorr},\ref{xdeltacorr}).
The 2D hole-hole correlation is plotted in Fig.~8a, which
shows a rapidly decaying central peak and a weak interference
pattern at larger distances due the presence of many-ionic impurities
and the d-wave gap structure. Taking the azimuthal average of Fig.~8a
results in Fig.~8b.  The decay length extracted from a Gaussian
fit is $\xi_x=1.8a$, very close to
that of the isolated ion case discussed in the previous section. 
This indicates that the main effects controlling the average size of
the ``patches'' come from the single ion screening cloud.
The weak oscillatory structures at larger
separations is a result of the interference pattern in Fig.~8a.
The distance to the first weak secondary peak can be interpreted
as the averaged distance between regions or patches where the
$x_i$ and $\Delta_d(i)$ vary significantly from their averaged
values.

The azimuthally averaged auto-correlation for the d-wave
order parameter variations is plotted in Fig.~8c which shows
the same structure with a decay length $\xi_{\Delta_d}\simeq 2.2a$.
We remark that at this average hole concentration, which is slightly
below optimal doping, the oscillatory structures are rather weak and are
just beginning to emerge. Interestingly, the azimuthally averaged correlation
functions obtained from the experimental STM data on optimally doped
BSCCO, shown in Fig.~2 (e-f) of Ref.~\cite{pan}, have the same structure, 
including the weak oscillations in the tails. We expect the
secondary peak to develop more prominently in more underdoped systems where
percolative-like patches are likely to become more pronounced with increasing 
average inter-hole distance.
%
\begin{figure}   
\center   
\centerline{\epsfxsize=3.4in   
\epsfbox{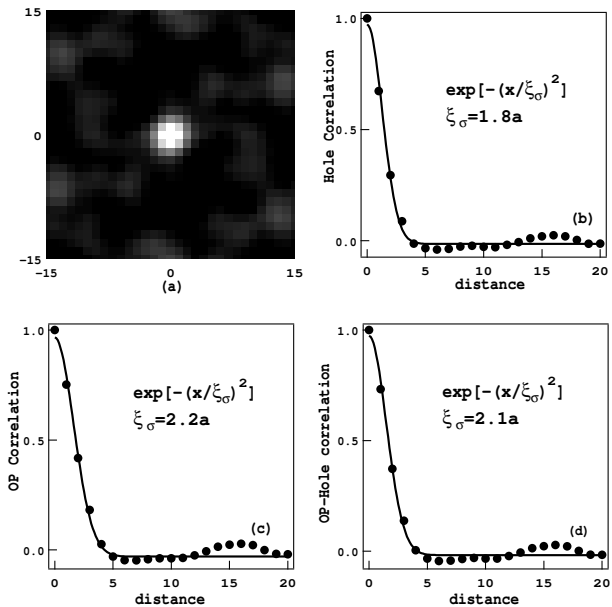}
}
\begin{minipage}[t]{8.1cm}   
\caption{The auto- and cross-correlation analysis of the spatial variations
in the doping concentration $\delta x_i$ and the d-wave order parameter
$\delta \Delta_d(i)$. (a) 2D auto-correlation of $\delta x_i$. (b)
Azimuthally averaged auto-correlation function of $\delta x_i$. (c) Azimuthally
averaged auto-correlation function of $\delta\Delta_d(i)$.
(d) The absolute value of the azimuthally averaged cross-correlation
function of $\delta x_i$ and $\delta\Delta_d(i)$. 
The solid lines in (b), (c) and (d) are Gaussian fits to the correlation
functions with the corresponding decay lengths $\xi_\s$ shown in legends.
} 
\label{correlations}    
\end{minipage}   
\end{figure}   

Comparison of the $x_i$-map in Fig.~7a and the $\Delta_d$-map
in Fig.~7b shows that the order parameter gap is bigger where
the doping rate is lower and vice versa, in agreement with 
the general trend of the doping dependence of $\Delta_d$ in a clean
system. This (anti-)correlated behavior is confirmed by the
cross-correlation analysis of $\delta x_i$ and $\delta\Delta_d$
in Fig.~8d, where the {\it absolute} value of the azimuthally averaged
correlation function is plotted versus separation. The fact that
the cross-correlation length, $\xi_{x-\Delta_d}\simeq2.1a$, 
is very close to that of the auto-correlations indicates
that the correlation between the spatial variations in 
$\Delta_d$ and $x_i$ is very strong.
\begin{figure}   
\center   
\centerline{\epsfxsize=2.6in   
\epsfbox{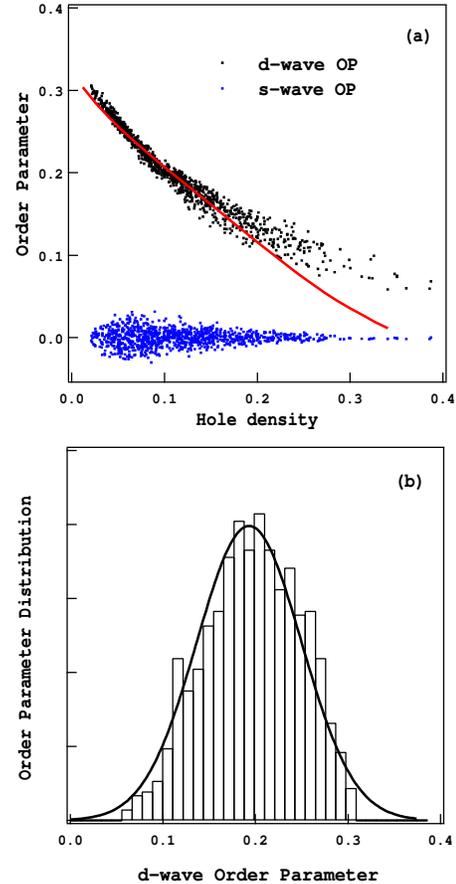}
}
\begin{minipage}[t]{8.1cm}   
\caption{ (a) A scatter-plot of the d-wave order parameter $\Delta_d(i)$
and the extended s-wave order parameter $\Delta_s(i)$ versus local 
doping concentration $x_i$. The solid line corresponds to $\Delta_d$
as a function of $x$ obtained in the absence of
the ionic dopant potential. (b) A histogram showing the statistical
distribution of $\Delta_d(i)$ in this {\it single} $32\times32$ system.
The solid line is a Gaussian fit.
} 
\label{opstat}    
\end{minipage}   
\end{figure}   

We proposed, in Ref.~\cite{pan}, the concept of {\it local doping 
concentration} to emphasize the local nature of the physics associated
with screening and pairing in a doped Mott insulator. To further
substantiate this idea, we produce a scatter-plot of the local d-wave order
parameter $\Delta_d(i)$ versus the local doping $x_i$ in Fig.~9a. 
For a range of doping near the 
average doping value of $x=0.12$, the data
points scatter around the $\Delta_d$ v.s. $x$ curve (solid line)
obtained in the homogeneous case without the 
ionic dopant potential. In the regions where the local doping
is low, the scatter in $\Delta_d$ is small and
the correlation between $\Delta_d(i)$ and $x_i$ is extremely
strong. On the other hand, the $\Delta_d(i)$ values show significant
scatter in the locally overdoped regimes and deviate appreciably
from the values in a clean system
at the corresponding doping level. 
Also shown in Fig.~9a is the scatter plot of the much smaller, extended
s-wave order parameter $\Delta_s(i)$ versus $x_i$. 
It scatters around a zero mean and shows a
slightly larger variance with decreasing LDC. The statistical
distribution of $\Delta_d$ in this single $32\times32$ system
is plotted as histograms in Figs.~9b. It can
be fitted reasonably well by a Gaussian distribution.

\subsection{Local tunneling density of states as measured by STM
spectroscopy}

To enable further comparisons with the STM experiments, we
next calculate the local tunneling density of states $N_i(\omega)$
according to Eq.~(\ref{ldos}). The $\delta$ functions in Eq.~(\ref{ldos})
are replaced by the derivatives of the corresponding 
Fermi distribution function\cite{allan}
at an inverse temperature of $\beta J=30$. Parallel to our presentation
of the experimental data\cite{pan}, we extract the local pairing energy
gap (full gap), denoted by $\Delta_T(i)$ to differentiate 
from the d-wave OP $\Delta_d(i)$ discussed above, 
from the peak-to-peak distance in the electron LDOS
spectrum $N_i(\omega)$
at every site $i$. Simultaneously, we obtain the integrated 
LDOS (ILDOS) at every site $i$ from
\eq
I_i=\int_{-\omega_0}^0 d\omega N_i(\omega),
\label{ildos}
\ee
where the integration limit is typically taken to be
$-\omega_0=-J$. Note that $N_i(\omega)$
at $\omega<0$ corresponds to the LDOS of the occupied states
below the chemical potential. It should be compared to
the STM tunneling spectrum at {\it negative sample bias} where electrons
tunnel out of the occupied states in the sample.
The energy $\omega$ is related to the bias $V$
according to $\omega=eV$.

The spatial variations of $\Delta_T(i)$
and $I_i$ are shown as the gap and the ILDOS maps
in Figs.~10a and 10b respectively. As used in Ref.~\cite{pan},
the reverse color coding in the gap map
(brighter color for a smaller tunneling gap)
makes it easy to observe its (anti-)correlation with the ILDOS map,
namely, bigger gap regions correspond to smaller ILDOS and
vice versa, in agreement with the STM data. In Fig.~11a, we
construct a scatter plot of the superconducting gap versus
the ILDOS consisting of a total of $32\times32=1024$ data points,
which suggests an approximate linear relationship between
the two quantities, in remarkable agreement with the STM data 
we presented in Fig.~4 of Ref.~\cite{pan}.

It is easy to notice the great similarity between the ILDOS map in Fig.~10b 
and the 2D hole density map shown in Fig.~7a.
This supports within our model the conjecture made in Ref.~\cite{pan}
that the ILDOS as measured by STM should be approximately proportional to 
the LDC.
\begin{figure}   
\center   
\centerline{\epsfxsize=3.0in   
\epsfbox{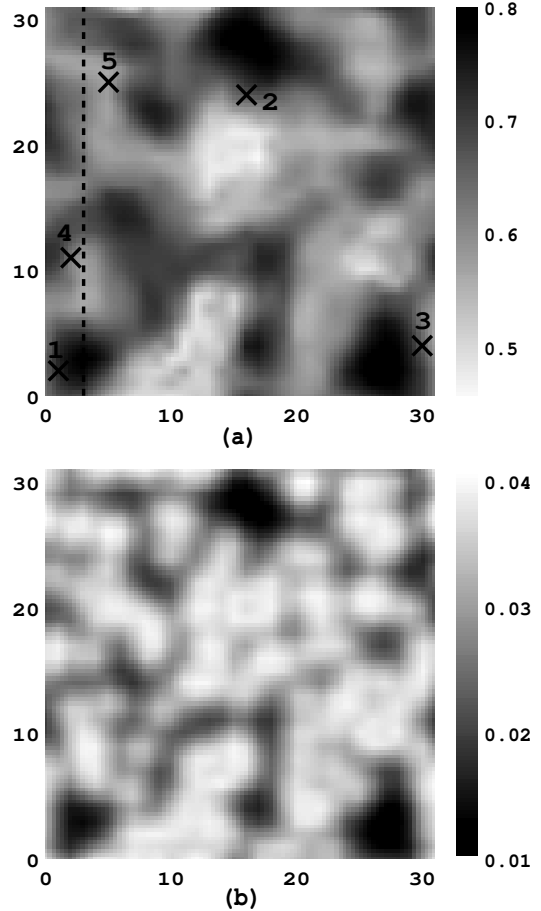}
}
\begin{minipage}[t]{8.1cm}   
\caption{The 2D tunneling gap map (a) and the ILDOS map (b).
The reverse color coding in the gap map
(brighter color for a smaller gap) 
makes it easy to observe its (anti-)correlation with the ILDOS map.
The tunneling spectra along 
a line-cut marked by the dashed line
and at the five locations marked by $\times$ in (a) are shown in Fig.~13.
} 
\label{gapildosmap}    
\end{minipage}   
\end{figure}   

To examine this relationship in more detail, we
show a scatter-plot of the ILDOS $I_i$, obtained from Eq.~(\ref{ildos}),
versus the local 
doping level $x_i$ in Fig.~11b. Remarkably, the ILDOS is indeed
linearly proportional to the LDC when
$x_i$ is small, i.e. when the local region is underdoped. 
This is consistent with the Mott-Hubbard picture 
in which doping a Mott insulator introduces spectral weight 
near the Fermi energy
necessary to transform the insulator into a superconductor.
As observed in recent ARPES experiments,\cite{shen,hong,ino}
the integrated quasiparticle weight scales with the
average doping concentration in the underdoped regime. The idea
of LDC that we advocate extends
this Mott-Hubbard picture, commonly used to interpret macroscopic
spectroscopy properties,
to microscopic scales. It remains valid in describing
the local spectroscopies of the short-coherence length superconducting
state of the doped Mott-insulator. 
However, for larger hole concentration, i.e. further away from the
Mott insulator, the ILDOS in Fig.~11b tends to tapper off and saturate.
It shows that the two quantities
are more intricately connected than the overall proportionality
stemming from the quasiparticle wave-function renormalization
which is apparent in Eq.~(\ref{ldos}). This result
turns out to be important for the analysis of the low energy
inhomogeneity structures in the following section.
\begin{figure}   
\center   
\centerline{\epsfxsize=3.0in   
\epsfbox{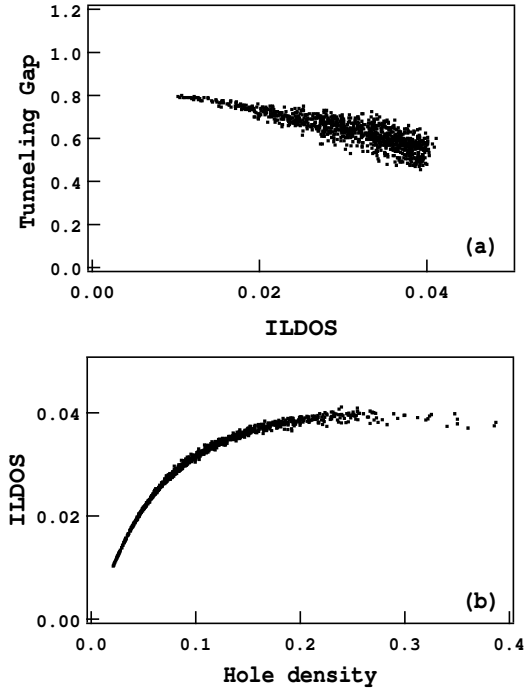}
}
\begin{minipage}[t]{8.1cm}   
\caption{(a) A scatter-plot of the superconducting gap $\Delta_T(i)$ in
the tunneling spectrum versus the ILDOS. (b) A scatter-plot of the
ILDOS versus the LDC $x_i$.
} 
\label{gapscatterplot}    
\end{minipage}   
\end{figure}

The statistical properties of the tunneling gap $\Delta_T$
and the ILDOS $I_i$ are in general quite similar to those of the 
d-wave pairing order parameter $\Delta_d$ and the local hole
concentration $x_i$. The histogram of $\Delta_T$ in the
$32\times32$ system is plotted in Fig.~12a, which shows
an approximate Gaussian distribution.
The azimuthally averaged auto-correlation of
the spatial gap variation $\delta\Delta_T(i)$ is shown in 
Fig.~12b, and that of the cross-correlation
of the gap variation and the ILDOS variation,
\eq
C_{I-\Delta_T}(\v r_j)={1\over N_s}\sum_i
\langle \delta \Delta_T(i) \delta I_{i+j}\rangle,
\label{ideltatcorr}
\ee 
is shown in Fig.~12c in the absolute value. 
The extracted gap correlation length $\xi_{\Delta_T}\simeq2.9a$ is somewhat 
larger than that of d-wave order parameter $\xi_{\Delta_d}$. 
Although the values of the decay lengths depend on the
value of the setback distance $d_s$ of the dopant oxygen ions, we
find for $d_s=1.5a$, and $a=3.8$\AA, $\xi_{\Delta_T}\simeq 11$\AA \ which
is close to the corresponding value of $14$\AA \ obtained by 
analyzing the experimental data.\cite{pan}
The similar cross-correlation length is consistent with the observation that 
these two quantities are strongly (anti-)correlated.
\begin{figure}   
\center   
\centerline{\epsfxsize=3.0in   
\epsfbox{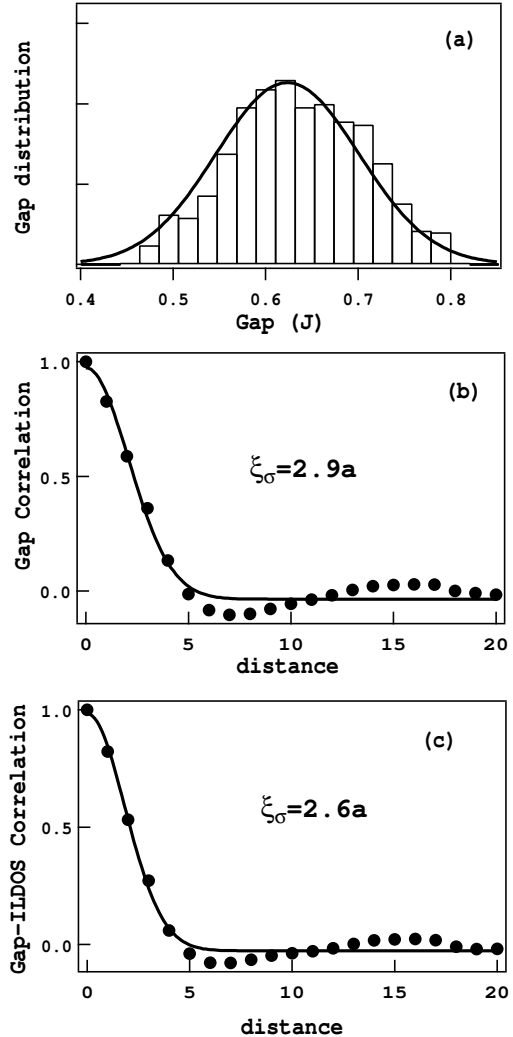}
}
\begin{minipage}[t]{8.1cm}   
\caption{Statistical properties of the tunneling gap $\Delta_T$.
(a) A histogram of $\Delta_T$ in the $32\times32$ system at
$x=0.12$, showing approximate Gaussian distribution (solid line).
(b) The azimuthally averaged auto-correlation function of the
tunneling gap variations.
(c) The absolute value of the azimuthally averaged cross-correlation
function between the variations in $\Delta_T$ and the ILDOS.
The solid lines are Gaussian fits.
} 
\label{gapstat}    
\end{minipage}   
\end{figure}

Next we turn to the details of the spatial variations of the
local tunneling spectrum and make comparisons to experiments.
Fig.~13a shows a three-dimensional rendering of the tunneling spectra
along a horizontal line-cut marked in Fig.~10(a), exemplifying 
the detailed variations of the
LDOS, the pairing energy gap, and the correlation between them.

\begin{figure}   
\center   
\centerline{\epsfxsize=2.6in   
\epsfbox{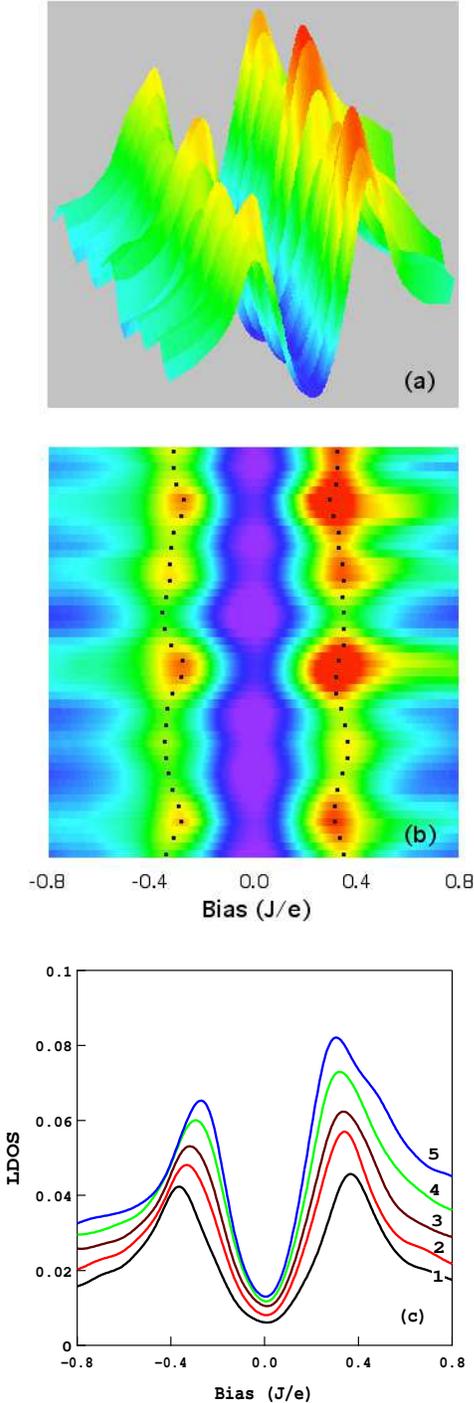}}
\centerline{\epsfxsize=2.6in 
\epsfbox{Fig11C_V6.EPSF}}
\begin{minipage}[t]{8.1cm}   
\caption{Spatial variations of the LDOS spectrum as a function
of bias. (a) A three-dimensional rendering of the tunneling spectra along
the line-cut indicated in Fig.~10(a). (b) The same data from a bird's-eye
view. The black dotted lines trace the locations of the coherence peak.
(c) Five characteristic spectra taken at the positions
marked in Fig.~10(a). The local doping concentrations are:
$x_1=0.05$, $x_2=0.07$, $x_3=0.10$, $x_4=0.14$, $x_5=0.19$.
} 
\label{linecut}    
\end{minipage}   
\end{figure}   
The projection of Fig.~13a onto the 2D plane is shown in Fig.~13b.
It is clear that regions with higher ILDOS also exhibit
smaller pairing gaps and higher coherence
peaks. The black dotted lines trace the positions
of the coherence peaks. Combined with the gap-map in Fig.~10a,
one can see that the gap varies somewhat less rapidly near the
center of a ``patch'' than at its ``edges''. We expect
this feature to become more prevalent for smaller average doping
concentrations, i.e. more underdoped samples. In Fig.~13c, we show 
five characteristic tunneling spectra calculated at the five
positions marked in Fig.~10a. The LDC
at these points are given in the captions. 
Notice that the spectral lines plotted versus the bias voltage
show remarkable resemblance to the
experimental data shown in Fig.~3 of Ref.~\cite{pan}. The most
striking feature is that the spatial inhomogeneity exists all
the way down to low energies as can be clearly seen from the
the systematic scatter in the zero-bias tunneling
conductance.\cite{notecurves12}
The broad range of agreements between the results presented in
Figs.~11, 12 and 13, and the experimental findings including
such details add to our confidence that the present theory of the 
inhomogeneous d-wave superconducting state with spinon pairing
and local holon condensation captures the essential physics 
of the superconducting state in BSCCO.

\section{Zero-bias tunneling density of states}

One of the most striking results of both the STM measurements
(see Fig.~3 of Ref.~\cite{pan})
and the present theory (see Fig.~13) is that the spatially inhomogeneous 
electronic structure not only appears at the energy scale of the 
superconducting gap, but persists all the way down to low energies,
including the zero-bias tunneling conductance.
On the experimental side, the low energy inhomogeneity can only
be observed clearly when the tunneling matrix element effect
along the $c$-axis is removed by a normalization
procedure introduced in Ref.~\cite{pan}.
Let's take a brief detour and study this
matrix element effect. The tunneling current at site ${\v r_i}$ is given
by
\eqa
I_E(\v r_iV)=\int d\omega &&f(\omega)[1-f(\omega+eV)]\times
\nonumber \\
&& N_i(\omega)
\rho(\omega+eV)\vert M(\v r_i,z)\vert^2,
\label{tunnelingcurrent}
\eea
where $\rho(\omega)$ is the density of states of the STM tip 
which is usually taken to
be a constant, and $M(\v r_i,z)$ is the tunneling matrix element.
Our knowledge of $M$ is scarce in complicated materials, but in general
one should be able to factor out its dependence on $z$ which corresponds
to the tip distance to the tunneling surface,
\eq
M(\v r_i,z)={\rm const.} \times m(\v r_i)e^{-\alpha\delta z(\v r_i)}.
\label{matrix}
\ee
Here $\delta z(\v r_i)$ is the spatial variation of the tip-to-surface
distance in the constant-current scanning mode, and $\alpha$ is a constant 
determined by the work-function. The detailed behavior
of $m(\v r_i)$ is presently unknown and can at best be assumed
to have no or very weak spatial dependence.  However, the spatial variation
in the tip distance $\delta z(\v r_i)$ is known in the form
of the surface topography from experiments.
The normalization procedure we used in Ref.~\cite{pan} is precisely
to remove this part of the matrix element effect such that the resulting
data represent tunneling measurements carried out with the tip-to-surface
distance effectively held constant. While it has been common practice
to present STM data without such a normalization, in which case the low
energy inhomogeneity of the tunneling conductance is not clearly 
revealed,\cite{howald,davis} we believe that the constant
tip distance normalization is
physical and should be carried out in order to obtain the true
electronic contribution to the tunneling spectra.
Our theoretical calculations
in the inhomogeneous d-wave superconducting state based
on {\it local} spinon pairing and {\it local} holon condensation
indeed support the experimental observation \cite{pan}
that the low energy STM tunneling
spectra is spatially inhomogeneous. We next provide a careful
analysis of the theoretical data at zero-bias and make a few predictions
that can be tested and used to guide further experiments.

Fig.~14a shows a 2D map of the zero-bias LDOS
at every point on our $32\times32$ lattice.
The spatial variations are obvious, but what is striking is that
this is the {\it same} pattern of inhomogeneity 
that we saw at the energy scale of the superconducting gap
as shown in the gap map in Fig.~10a, in complete agreement with
experiments.\cite{pan2} Moreover, comparing Fig.~14a
to the local hole density map in Fig.~7a shows that they are
directly correlated. This strongly suggests that the same kind
of inhomogeneity pattern should be observable at {\it all} energy scales
in the STM tunneling spectroscopy. To further illustrate the correlation
between them, we construct a scatter-plot of the
zero-bias LDOS $N_i(0)$ versus the LDC $x_i$ in Fig.~14b. 
It suggests a strong correlation via a linear relationship 
between the two quantities in the regions of low doping. The scatter
increases gradually with increasing LDC and
the linear relationship tappers off and saturates. This is quite
similar to the behavior of the ILDOS and points to the more intricate
$x_i$-dependence of the LDOS in Eq.~(\ref{ldos}) 
beyond the overall linear proportionality.

Since the local doping concentration is not
directly accessible experimentally, in Fig.~14c we present the scatter
plot of the zero-bias LDOS versus the ILDOS. Interestingly, 
due to the weaker variation of
the ILDOS in regions of higher hole concentration (see Fig.~14b),
the scatter of the zero-bias LDOS exhibits a slight upward curvature
with increasing ILDOS. Such a scatter plot
can be directly constructed from the STM experimental data for
comparison,\cite{pan2} which serves as another test for the present theory.
Two remarks are in order. 

(i) It is important to emphasize the difference between the electron
tunneling DOS that spectroscopy measurements probe and the
thermodynamic DOS or the compressibility that shows up in thermodynamic
measurements such as the specific heat.  
While the former depends on both the 
quasiparticle wave-function renormalization $Z$ and the self-energy
corrections, the latter is insensitive to $Z$.\cite{elihu}
Although the inhomogeneity in the tunneling spectra
at low energies emerges from the spatial variations in
the LDC $x_i$ through both the wave-function renormalization 
and the spinon tunneling spectra as described by Eq.~(\ref{ldos}), 
its implications on the thermodynamic DOS
and the transport properties require
a different analysis. We stress that the difference between these two 
density of states even at our unrestricted mean-field level 
is more than the local doping concentration $x_i$, the
prefactor in the electron tunneling LDOS,
due to the presence of the spinon wave-functions in Eq.~(\ref{ldos}).

\begin{figure}   
\center   
\centerline{\epsfxsize=3.0in   
\epsfbox{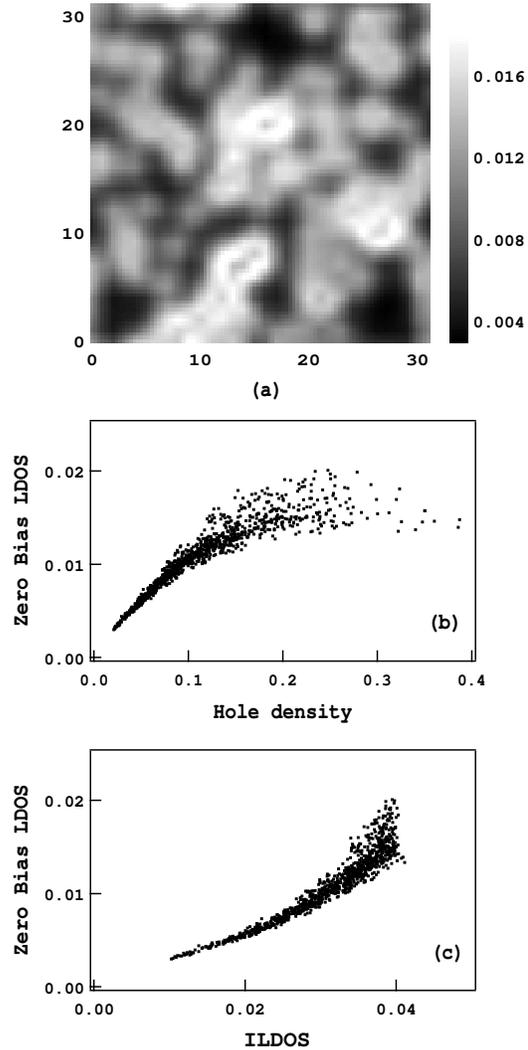}
}
\begin{minipage}[t]{8.1cm}   
\caption{Spatial inhomogeneity at zero bias. (a) A 2D map of the
zero bias LDOS, $N_i(0)$, showing the same pattern of inhomogeneity
as the tunneling gap map and the ILDOS map in Fig.~10. (b) A scatter-plot
of $N_i(0)$ versus the LDC $x_i$. (c) A scatter-plot
of $N_i(0)$ versus the ILDOS.
} 
\label{zbldos}    
\end{minipage}   
\end{figure}

(ii) Although the spinon tunneling DOS is not directly
accessible by STM measurements, since only the physical electrons
can tunnel in and out of the sample, it can be, nevertheless,
readily extracted from Eq.~(\ref{ldos}). Dividing out the prefactor
$x_i$, we show in Fig.~15a the spinon tunneling spectra,
$N_i^f(\omega)=N_i(\omega)/x_i$, at the five locations marked in Fig.~10a.
Comparing to the corresponding spectral lines for electron LDOS
at the same locations shown in Fig.~13c, it is clear that 
the degree of inhomogeneity remains large 
at the energy scale of the superconducting gap. This is not
surprising because the gap inhomogeneity results from
local spinon pairing in our picture. However, the degree of
inhomogeneity appears to have been somewhat reduced at low energies
near zero-bias, suggesting that the spinons near the gap nodes, unlike the
(electron) nodal quasiparticles, experience less 
inhomogeneity.\cite{dhlee2}
The scatter plot of the spinon zero-bias LDOS versus local hole
concentration is shown Fig.~15b. Although the relative magnitude of
the variations is reduced by about $30\%$ 
from the electron case in Fig.~14a, the spatial variation 
of the zero-bias conductance for spinons is not only clearly visible,
but shows a very well defined correlation with the local hole concentration
that is remarkably similar to that of the d-wave order parameter $\Delta_d$
shown in Fig.~9a. Therefore, we conclude that the low energy spinons
near the d-wave gap nodes will experience the same type of 
inhomogeneity as the spinons at the gap edge.
\begin{figure}   
\center   
\centerline{\epsfxsize=3.6in   
\epsfbox{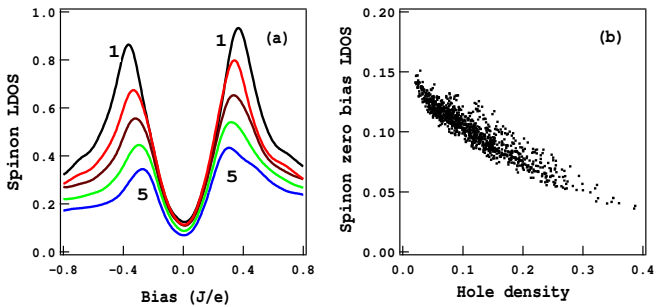}
}
\begin{minipage}[t]{8.1cm}   
\caption{The tunneling LDOS for spinons. (a) The spinon tunneling spectra
at the five positions marked in Fig.~10a, in comparison to
those of the electron tunneling spectra shown in Fig.~13c where the
LDC $x_i$, $i=1,\dots,5$  were given.
(b) A scatter-plot of the spinon zero-bias LDOS versus
the LDC $x_i$, showing
remarkable similarity to that of $\Delta_d(i)$ versus $x_i$ presented in
Fig.~9a.
} 
\label{spinonldos}    
\end{minipage}   
\end{figure}

\section{STM topography and the tunneling spectra at constant
current}

We next derive from our numerical data, a theoretical STM topographic image
at constant tunneling current for the $32\times32$ system we study. We then
use the result to reconstruct the local tunneling spectra at constant
current, which corresponds to the original STM data before normalization.
In doing so, we will further illustrate the physics behind the
normalization procedure used in Ref.~\cite{pan}, which is essentially 
a mapping from a constant current topography
to the electron LDOS measured at constant tunneling barrier width. 

The STM measurements are usually carried
out at a constant tunneling current $I_0$ which is equivalent to
the integrated LDOS from the sample bias voltage $-V_0$ to the Fermi level. 
Thus we can write,
\eq
I_0=\int_{-V_0}^0 dV {dI\over dV}(\v r,V,z(\v r)),
\label{i0}
\ee
where ${dI\over dV}(\v r,V,z(\v r))$ is the LDOS or the 
tunneling differential conductance. It is a function of the 2D 
coordinates ${\v r}$ on the tunneling surface, the sample bias $V$, 
and the tip to surface distance $z(\v r)$ (topography) that must vary 
at every ${\v r}$ in order to keep the current $I_0$ a constant. 
Note that ${dI\over dV}$ in Eq.~(\ref{i0})
corresponds precisely to the original STM data before normalization. 
From our discussion following Eq.~(\ref{tunnelingcurrent}), it is 
clear that ${dI\over dV}$ is related to the
physical electronic tunneling spectra $N(\v r,eV)$ obtained
at constant $z(\v r)$ according to,
\eq
N(\v r,eV)={dI\over dV}(\v r,V, z(\v r)) e^{\alpha z(\v r)}.
\label{mapping1}
\ee
Integrating Eq.~(\ref{mapping1}) over $V$ from $-V_0$ to
$0$ gives, after using Eq.~(\ref{i0}),
\eq
I(\v r)= I_0 e^{\alpha z(\v r)},
\label{mapping2}
\ee
where $I(\v r)$ is the electronic ILDOS given by Eq.~(\ref{ildos})
in terms of the normalized or the calculated tunneling spectra.
Eq.~(\ref{mapping2}), together with Eq.~(\ref{mapping1}), completely describes
the mapping between the constant current topography and the local electron
tunneling spectra at constant tunneling barrier width.

From our calculated ILDOS map and using Eq.~(\ref{mapping2}), we obtain
the constant current STM topographic image $z(\v r)$ on our
$32\times32$ lattice using $\alpha=2.3$\AA$^{-1}$ which corresponds
to a work-function of about $4$eV. The topography is presented
in Fig.~16a, which exhibits
identical structures as the electron ILDOS map shown in Fig.~10b.
We next covert the entire local tunneling spectra $N(\v r,V)$
at constant tip-sample distance to the tunneling differential conductance
at constant current using Eq.~(\ref{mapping1}) and the topography map.
The procedure turns out to be equivalent to normalizing the 
LDOS spectra by the corresponding ILDOS.
\begin{figure}   
\center   
\centerline{\epsfxsize=3.0in   
\epsfbox{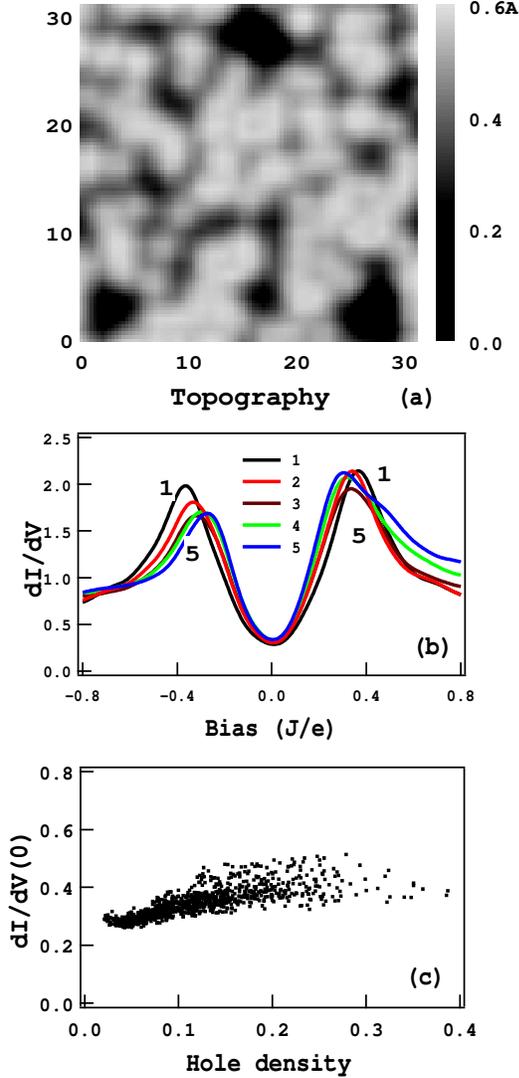}
}
\begin{minipage}[t]{8.1cm}   
\caption{Calculated STM spectroscopy of the $32\times32$ lattice
in the constant current mode. (a) The topographic image of our system,
showing the same inhomogeneity
structures as the electron ILDOS map plotted in Fig.~10b.
(b) The characteristic tunneling differential conductance spectra
at the same five positions marked in Fig.~10a. The integrated area under
the spectral lines on the negative bias side is the same, i.e. the
constant tunneling current. Note that the inhomogeneous distribution
of tunneling differential conductance near zero bias is hardly visible.
(c) A scatter-plot of the zero-bias conductance versus the LDC.
} 
\label{topography}    
\end{minipage}   
\end{figure}   

In Fig.~16b, we show five differential conductance curves
at the same five positions marked in Fig.~10a.
Notice that, typical of the raw STM data,
the spectral lines on the negative bias side cross
approximately at one value of the voltage as a result of the constraint
that the integrated area under each curve, i.e. the tunneling
current, must be the same. The spatial inhomogeneity on the energy
scale of the superconducting gap remains, albeit that the peak height
does not show the same systematics --- 
smaller gap with higher coherence peak --- as in the experimental data 
with or without normalization. \cite{pan2}
This discrepancy may be due to the simplification of
the mean-field theory which ignores the effects of fluctuations, and
perhaps also, to a certain extent, due to the complication of the matrix 
element effects in the STM tunneling data. The spatial variations
of the tunneling differential conductance spectra
at low energies are, however, much less visible.
The scatter-plot of the zero-bias conductance
versus local doping concentration is shown in Fig.~16c. Comparing to
the same plot for the electron zero-bias LDOS shown in Fig.~14b, we find
that the relative magnitude of the variations in the zero-bias conductance
is suppressed by about $65\%$
and it shows almost no systematic dependence on the LDC.
These findings are in good qualitative agreement with the experimental
observation that the original STM data before normalization show much less
spatial inhomogeneity at low energies.

\section{Summary and Discussions}
 
We have investigated in this paper, through microscopic calculations based on 
a generalized t-J model, the effects of the off-plane ionic potential 
associated with off-stoichiometry doping in the d-wave superconducting state
of a doped Mott insulator.
We find that nonlinear screening dominates the
response of the doped Mott insulator to the ionic potential.
One of the main characteristics of nonlinear screening is the emergence
of a percolative-type of inhomogeneity in the electronic
structure. In 2D electron systems
formed in modulation doped semiconductors, 
nonlinear screening of the ionic dopant potential at low electron densities
can tear the 2D electrons into an inhomogeneous mixture of metallic and 
dielectric regions.\cite{efros} What we have shown here
is a striking analogy of the physics in a doped Mott insulator: 
nonlinear screening of the dopant ionic potential leads to an 
inhomogeneous d-wave superconducting state wherein the LDC
and the d-wave superconducting gap exhibit 
significant spatially correlated variations on the scale of
a nanometer comparable to the
short coherence length. In this context, the concept of a
{\it local doping concentration} augmented by the generalization of
the Mott-Hubbard picture to local spectroscopies that we advocate \cite{pan} 
is very useful.  We have shown that, within our self-consistent, spatially
unrestricted mean-field approach, local spinon pairing and local 
holon condensation capture the essential physics of
the inhomogeneous superconducting state and provide remarkably
consistent descriptions of the experimental data.\cite{pan}

We have shown that there is one length scale controlled by the ionic
setback distance $d_s$ that characterizes the inhomogeneity.
The decay lengths of the auto-correlations of the spatial variations in 
the LDC and
the d-wave pairing order parameter are found to be close to $d_s$,
suggesting that the superconducting pair-size is determined by $d_s$.
In this paper, we have chosen $d_s=1.5a$ based on the physical
distance of $\sim 5$\AA \ between the CuO$_2$ plane and the BiO 
layer where the dopant oxygen ions reside and $a\simeq3.8$\AA \ 
for the Cu-Cu atomic spacing. However, it is important to emphasize
the effects of anisotropy in the background dielectric screening constant
which enters the ionic potential.
It was pointed out to us by Kivelson
that, for the high-T$_c$ cuprates, the background dielectric constant,
determined by the electronic interband polarization and phonon contributions,
is highly anisotropic, i.e., it has a different value along the 
$c$-direction, $\epsilon_\perp$, than in the $ab$-plane $\epsilon_\parallel$.
If this anisotropy is taken into account, the ionic potential
given in Eq.~(\ref{ionpotential}) must be modified by 
replacing\cite{kivelsonprivate}
$V_d\to V_d^*$ and $d_s\to d_s^*$  with $d_s^*=\sqrt{\epsilon_\parallel
\epsilon_\perp}d_s$. Thus the {\it effective} distance between
the BiO layer and the CuO$_2$ plane can be significantly larger
than $5$\AA. Taking $\epsilon_\perp\simeq8.0$ and $\epsilon_\parallel\simeq
1.5$, it follows that the effective setback distance 
$d_s^*\simeq3.5a\simeq 13$\AA. Our results then imply that the
pair-size, determined by the decay length in the spatial
variation of the d-wave order parameter, would be on the order of one
to two nanometers, in good agreement with our experimental findings.\cite{pan}

The physics discussed here can be continued to more 
underdoped cases where the percolative structures resulting from 
nonlinear screening are expected to be more pronounced. 
Already in the case of $12\%$ average doping studied here, 
the tunneling spectra along particular line-cuts in Fig.~10a 
show patches over which the superconducting gap is nearly flat 
in the center and changes quickly near the edges,
indicative of a percolative electronic structure which is often
referred to as microscopic phase separation.\cite{kivelson,howald,davis}
As we have shown that the size of the screening cloud around
an isolated ion is determined by its effective setback distance $d_s^*$, 
it is likely that, with increasing doping, a percolation transition/crossover
takes place near optimal doping where
the average inter-hole distance becomes comparable to
an effective setback distance $d_s^*$.

We have not discussed the very underdoped
physics in detail because in that case it will
become important to include the effects associated with magnetism that 
is not important and
has thus been left out in the average doping range considered in this paper. 
We have tested that if magnetism is
included, locally antiferromagnetic ordered insulating regions 
emerge and percolate at sufficiently low averaged 
doping. This is left for further studies. Nevertheless, the present
theory implies that in underdoped BSCCO, the local tunneling spectra
should reveal a percolative structure of superconducting
patches embedded in a background where larger tunneling gaps to
single-particle excitations arise.

A frequently asked question is how does the high transition temperature 
coexist with the inhomogeneity. While we do not have a complete
answer, we would like
to discuss a few relevant issues related to this question.
One of the most obvious reasons for high-$T_c$ 
superconductivity to survive
the inhomogeneity is its short coherence length.
In our theory, the latter, determined by the
effective setback distance $d_s^*$, is also the scale of
the inhomogeneity and will therefore remain shorter than the 
mean-free path which is usually much larger than the correlation
length of the scattering centers for a reasonable scattering strength.
Moreover, as we have pointed out in Ref.~\cite{pan}, the
scattering by the out-of-plane ionic potential involves predominantly
small momentum transfers limited by the
small scattering angle on the order of ${1/d_s^* k_F}$, where 
$k_F$ is the Fermi momentum. This type of scattering is much less
effective at reducing $T_c$ than at increasing the single-particle
scattering rate.\cite{kee} 
\begin{figure}   
\center   
\centerline{\epsfxsize=2.6in   
\epsfbox{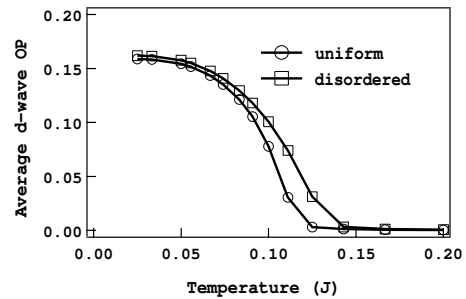}
}
\begin{minipage}[t]{8.1cm}   
\caption{The temperature dependence of the {\it averaged} d-wave
order parameter (open squares) on a $16\times16$ lattice at 
an average doping $x=0.15$. Also shown in open circles is the
temperature dependence of the d-wave order parameter in the
uniform d-wave state without the ionic potential.
} 
\label{Tc}    
\end{minipage}   
\end{figure}   
We have preliminary results on the temperature dependence of the distribution
of the d-wave order parameter $\Delta_d$. 
In Fig.~17, we plot the {\it averaged}
$\Delta_d$ versus temperature on a $16\times16$ lattice at
an average doping $x=0.15$. Also shown is the temperature
dependence of $\Delta_d$ in the same system without the ionic
potential. Comparing the two curves shows no sign of degradation
of $T_c$ defined by the averaged order parameter.
Nevertheless, a distribution of $\Delta_d$ does imply a distribution of 
$T_c$ and a finite width for the transition.
We leave this topic for future study.

There is no evidence at present that would suggest the inhomogeneity
as being necessary for observing high-$T_c$ superconductivity. However,
the existence of such inhomogeneity should at least help protect
the d-wave superconducting state from other possible 
competing instabilities. Similarly, from a theoretical perspective,
the inhomogeneous d-wave superconducting state with spinon pairing
and local holon condensation may do much better
at suppressing low-lying fluctuations beyond the self-consistent
mean-field solutions than its uniform counterpart.

We next briefly discuss what happens if the dopant ions are 
more ordered in the BiO layer. Experimentally, 
thermal annealing is believed to be able to achieve a certain
degree of dopant ordering. To this end, we studied a case where
the dopant ions are periodically ordered and the average doping
$x=0.14$. We found that the LDC and the d-wave
order parameter show spatial variations that are periodic with
the same periodicity as that of the dopant ions. 
The widths of the distribution functions are
significantly reduced when compared to the disordered case.
We therefore expect that, although the spatial inhomogeneity
would remain after thermal annealing, the spatial variations would
be more periodic in structure with a noticeably smaller magnitude.

Based on these results, we expect that the microscopic variations
in the electronic properties of YBCO, particularly in its
ortho-phase,\cite{liang} should be more periodic in space
because the oxygen dopants can be ordered in the copper-oxygen chains.
However, there exists an important difference in the role of
the dopant ions. In BSCCO, the BiO plane is essentially insulating
which allows us to treat the dopants as solely
providing the dielectrically screened ionic potential.
The situation can be very different in YBCO because the chains themselves
are conducting and have low energy dynamics of their own.
The electrons in the chains will scatter off the dopant ions
giving rise to charge density oscillations along the chains that
may complicate the way by which carriers are doped into the
CuO$_2$ plane. The present theory would predict the appearance of
inhomogeneous electronic structures in the CuO$_2$
planes of YBCO that are most pronounced directly under/above the doped chains.
This type of spatial inhomogeneity on YBCO may have
been recently observed.\cite{alex}

We have investigated here one of the most direct consequences
of off-stoichiometry doping of a Mott insulator, 
namely the ionic potential. There are 
potentially others. For example, the presence of dopants can affect the
transfer integrals in their vicinity, causing spatial variations
in the parameters t and J, and perhaps more importantly in $t_\perp$ along
the $c$-axis. It is therefore plausible that certain aspects
of the inhomogeneous superconducting state could also arise from 
inhomogeneous pairing interactions\cite{larkin,simon} or 
impurities and defects in the superconducting plane\cite{atkinson,mohit}
and many other kinds studied previously in the context of 
disordered superconductors.
These are interesting and important issues that need to
be further investigated.

\section{Acknowledgments}

The authors thank Sasha Balatsky, Seamus Davis, Misha Fogler, Jung-Hoon Han, 
Hae-Young Kee, Steve Kivelson,
Dung-Hai Lee, Patrick Lee, Alex de Lozanne, and Allan MacDonald for 
useful discussions and comments. 
They are especially grateful to Steve Kivelson, Dung-Hai Lee,
and Patrick Lee for sharing their insights and to Jung-Hoon Han for 
his involvement and contributions during the early stage of this work.
They thank Jared O'Neal and Robert Badzey for their help with data
analysis. Z.W. would like to thank Patrick Lee for hosting his 
sabbatical leave at MIT where a part of this work was completed. 
This work is supported in part
by DOE Grant No. DE-FG02-99ER45747, NSF Grant No. DMR-0072205, 
and by the Sloan Foundation and Research Corporation.
The authors also thank the Institute for Theoretical Physics 
at UCSB where this
work was begun during the workshop on High Temperature Superconductivity
and acknowledge the generous support of NSF Grant No. PHY94-07194.

\vspace*{\fill}{

}

\begin{references}
\bibitem{note1} We find that in the presence of the dopant ionic
potential, the tendency towards phase separation is suppressed
even without the long-range Coulomb interaction in the doping range
of interest here.
\bibitem{notecurves12} We note that in Fig.~3 of Ref.~\cite{pan},
the two tunneling spectra taken at locations where the 
ILDOS is very small do not show the coherence peaks and the
superconducting gaps are absent indicative of insulating behavior.
The tunneling spectra in these rare insulating regimes
cannot be described by our present mean-field theory which does not include
the incoherent contributions that dominates the insulating behavior.
\end{references}
\end{document}